\documentclass[]{iau}

\usepackage{amsmath}
\usepackage{graphicx}
\usepackage{multirow}

\newcommand{\feh}{[Fe/H]}
\newcommand{\zmax}{$\langle Z_{\rm{max}} \rangle$}
\newcommand{\eccentricity}{$\langle e \rangle$}

\def\apjs{ApJS}

\def\aap{A\&A}

\def\mnras{MNRAS}

\begin{document}

\lefttitle{Dantas et al.}
\righttitle{The interplay between super-metallicity, lithium depletion, and radial migration in nearby stars}

\jnlPage{1}{7}
\jnlDoiYr{2025}
\doival{10.1017/xxxxx}
\volno{395}
\pubYr{2025}
\journaltitle{Stellar populations in the Milky Way and beyond}

\aopheadtitle{Proceedings of the IAU Symposium}
\editors{J. Mel\'endez,  C. Chiappini, R. Schiavon \& M. Trevisan, eds.}

\title{The interplay between super-metallicity, lithium depletion, and radial migration in nearby stars}

\author{M. L. L. Dantas$^{1\star,2,3}$, R.~Smiljanic$^{3}$, R. Boesso$^{4,5}$, H. Rocha-Pinto$^{5}$, L.~Magrini$^{6}$, G.~Guiglion$^{7,8,9}$, D.~Romano$^{10}$}
\affiliation{
$^{1}$Instituto de Astrofísica, Pontificia Universidad Católica de Chile, Av. Vicuña Mackenna 4860, Santiago, Chile\\
$^{2}$Centro de Astro-Ingeniería, Pontificia Universidad Católica de Chile, Av. Vicuña Mackenna 4860, Santiago, Chile\\
$^{3}$Nicolaus Copernicus Astronomical Center, Polish Academy of Sciences, ul. Bartycka 18, 00-716, Warsaw, Poland \\
$^{4}$Fundação Getulio Vargas’ Brazilian Institute of Economics (FGV IBRE), Rio de Janeiro, Brazil\\
$^{5}$Observat\'orio do Valongo, Universidade Federal do Rio de Janeiro, Ladeira Pedro Ant\^onio 41, 20080-090 Rio de Janeiro, Brazil\\
$^{6}$INAF -- Osservatorio Astrofisico di Arcetri, Largo E. Fermi, 5. 50125 Firenze, Italy\\
$^{7}$Zentrum f\"ur Astronomie der Universit\"at Heidelberg, Landessternwarte, K\"onigstuhl 12, 69117 Heidelberg, Germany\\
$^{8}$Max-Planck Institut f\"{u}r Astronomie, K\"{o}nigstuhl 17, 69117 Heidelberg, Germany \\
$^{9}$Leibniz-Institut f\"ur Astrophysik Potsdam (AIP), Potsdam, Germany\\
$^{10}$INAF -- Osservatorio di Astrofisica e Scienza dello Spazio, Via Gobetti 93/3, 40129 Bologna, Italy\\ 
$\star$ mlldantas@protonmail.com}

\begin{abstract}
We report the discovery of a peculiar set of old super-metal-rich dwarf stars with orbits of low eccentricity that reach a maximum height from the Galactic plane between $\sim$ 0.5-1.5 kpc observed by the \emph{Gaia}-ESO Survey. These stars show lithium (Li) depletion, which is anti-correlated with their [Fe/H]. To investigate these stars' chemo-dynamical properties, we used data from the \emph{Gaia}-ESO Survey. We applied hierarchical clustering to group the stars based on their abundances (excluding Li). Orbits were integrated using \emph{Gaia} astrometry and radial velocities from \emph{Gaia}-ESO. Our analysis suggests that the high metallicity of these stars is incompatible with their formation in the solar neighbourhood. We also found that their Li envelope abundance is below the benchmark meteoritic value, in agreement with previous works. This result supports the idea that the Li abundance in old, super-metal-rich dwarf stars should not be considered a proxy for the local interstellar medium Li.
\end{abstract}

\begin{keywords}
    Galaxy: kinematics and dynamics --
    Galaxy: abundances --
    Galaxy: stellar content
\end{keywords}

\maketitle

\section{Introduction}

This contribution presents our work on metal-rich and super-metal-rich (SMR) stars that have likely migrated from the inner Milky Way (MW) disc. We investigated the implications of this migration for lithium (Li) abundances in these stars. For more details, we refer the reader to \citet{Dantas2022, Dantas2023}, where this investigation is described in larger depth.

\section{Data \& Methodology}

We analysed a sample of 1460 stars located in the Galactic disc. Based on the concept of chemical enrichment flow presented by \citet{BoessoRochaPinto2018}, we employed hierarchical clustering (HC), stratifying the sample into six groups using 21 species of 18 elements. Our focus was on five SMR subgroups, comprising a total of 171 stars. Additionally, we extended the analysis to include groups with super-solar metallicities to investigate the relationship between radial migration and lithium (Li) depletion. Stellar ages were determined using \textsc{unidam} \citep{Mints2017}, yielding median ages of $\sim$ 8 Gyr. Orbital integrations were performed with \textsc{galpy} \citep{Bovy2015}, using \emph{Gaia} parallaxes and radial velocities from the analysis of spectra obtained with the UVES spectrograph at the Very Large Telescope ($R \sim 47000$) as part of the \emph{Gaia}-ESO Survey \citep[iDR6;][]{Gilmore2022, Randich2022}.

\section{Evidence of migration}

\begin{figure*}
    \centering
    \includegraphics[width=\linewidth, trim={0 0.4cm 0 0}, clip]{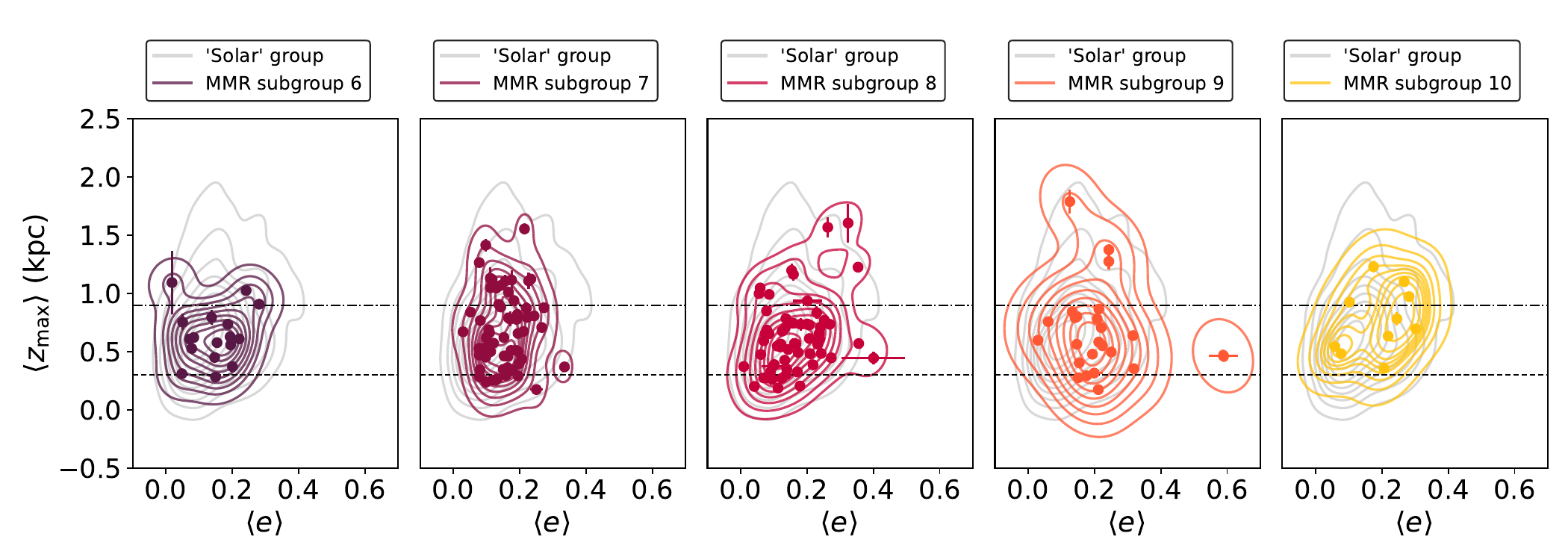}
    \includegraphics[width=\linewidth, trim={0 0.5cm 0 2.69cm}, clip]{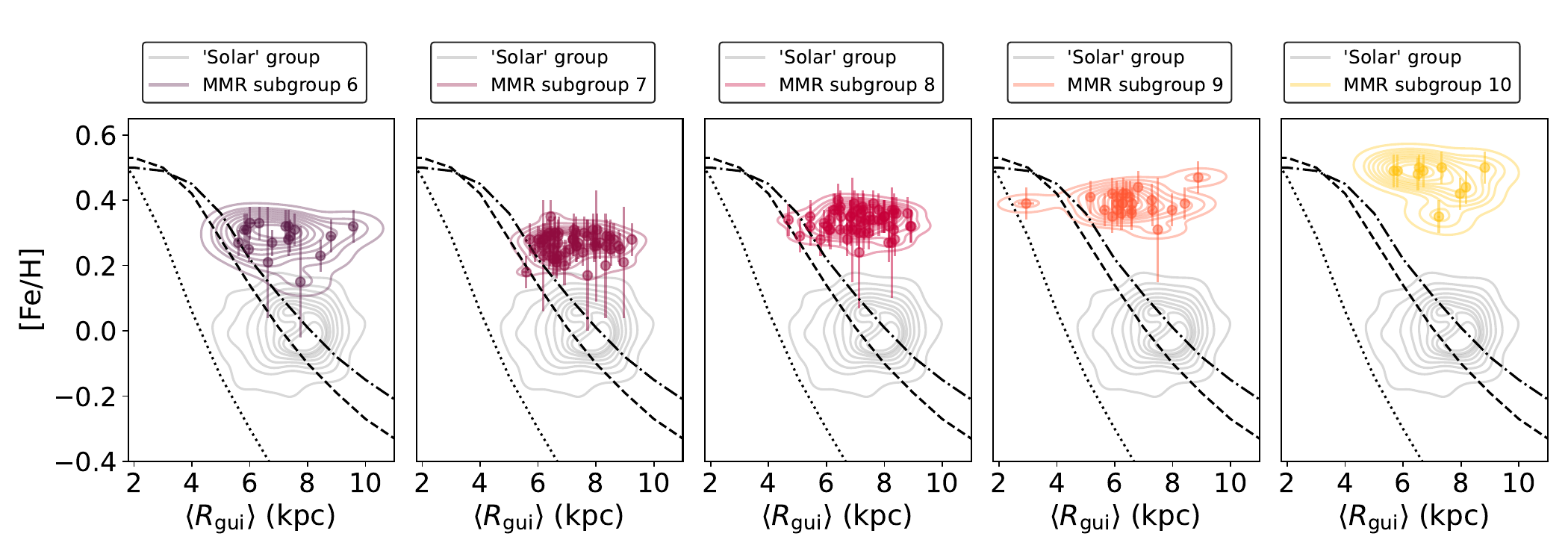}
    \caption{\emph{Top panel}: \zmax\ \emph{vs.} \eccentricity\ for the five subgroups of the most metal-rich group. The Solar-metallicity group parameters are shown in grey for comparison. The thin and thick disc scale-heights are shown as black dashed and dot-dashed lines respectively. \emph{Bottom panel}: similar to the top panel, but showing [Fe/H] \emph{vs.} $\langle R_{\rm gui} \rangle$ (guiding radius). The dotted, dashed, and dot-dashed black curves respectively depict the 3.3, 8, and 11 Gyr models described in \citet{Magrini2009}. The analysis encompassing the full sample with all metallicities is discussed in Dantas et al. (in review).}
    \label{fig:zmax_ecc}
\end{figure*}

Two key parameters for analysing whether a star has probably undergone churning (a change of its angular momentum/guiding radius due to dynamical interactions with the Galactic bar/spiral arms) are the maximum Galactic height (\zmax) reached during the stellar orbit and eccentricity (\eccentricity; see Fig. \ref{fig:zmax_ecc} top panel). Stars with a high probability of having undergone outward churning tend to reach substantial \zmax\ values \citep[e.g.][]{Roskar2013} and settle into more circular orbits \citep[i.e. small \eccentricity; e.g.][]{Khoperskov2020}. We observe that stars in all subgroups can indeed reach high \zmax, often exceeding the scale-height of the thick disc. Furthermore, as shown in Fig. \ref{fig:zmax_ecc} (bottom panel), all stars from the super-metal-rich subgroups exhibit much higher current guiding radii ($R_{\rm gui}$) than expected, as compared to the dotted curve, providing strong evidence for outward churning.

\section{Li depletion in metal-rich migrating stars}

Lithium is both produced and easily depleted in stellar photospheres. Radial migration has been linked to the observed decline in Li abundances among older, metal-rich dwarf stars currently found in the solar neighbourhood in certain surveys \citep[e.g.][]{Guiglion2019}. We investigated whether radial migration influences Li abundance patterns in dwarf stars in the solar neighbourhood and whether Li abundances in these stars reliably trace interstellar medium (ISM) Li levels. Using the aforementioned high-quality \emph{Gaia}-ESO data, we analysed both measured Li abundances and upper limits. The Li envelope of radially migrated stars falls below the benchmark meteoritic value ($<3.26$ dex), with the maximum observed abundance at A(Li) = 2.76 dex. This is consistent with previous findings for old dwarf stars (median age $\sim$8 Gyr), where Li decreases as [Fe/H] increases from the solar value. Our results suggest that Li abundance in these stars should not be used as a proxy for ISM Li levels and, thus, can not be used to constrain Galactic Li evolution models \citep[e.g. ][]{Romano2021}. Most importantly, these findings indicate that Li depletion in metal-rich dwarf stars can result from a combination of stellar evolution and radial migration.

\begin{figure}
    \centering
    \includegraphics[width=0.49\linewidth]{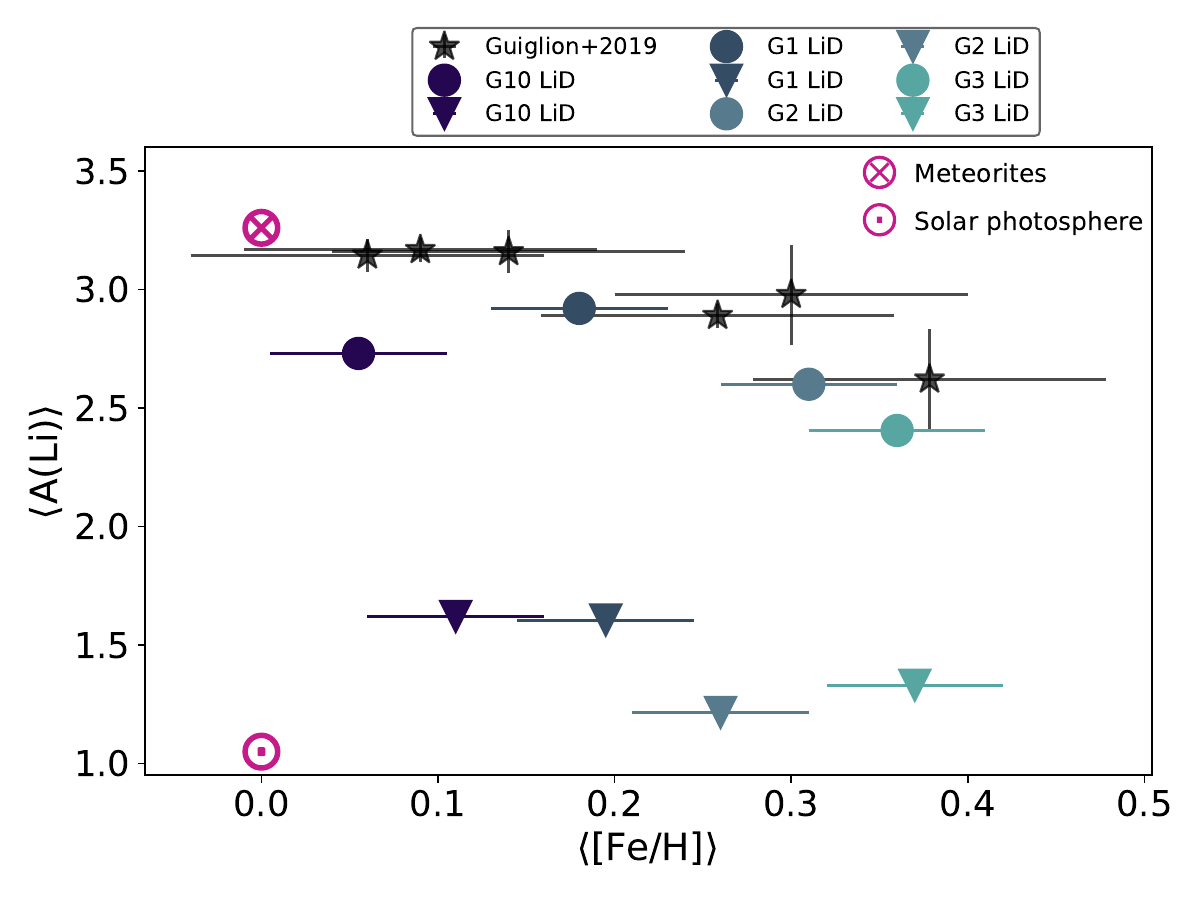}
    \includegraphics[width=0.49\linewidth]{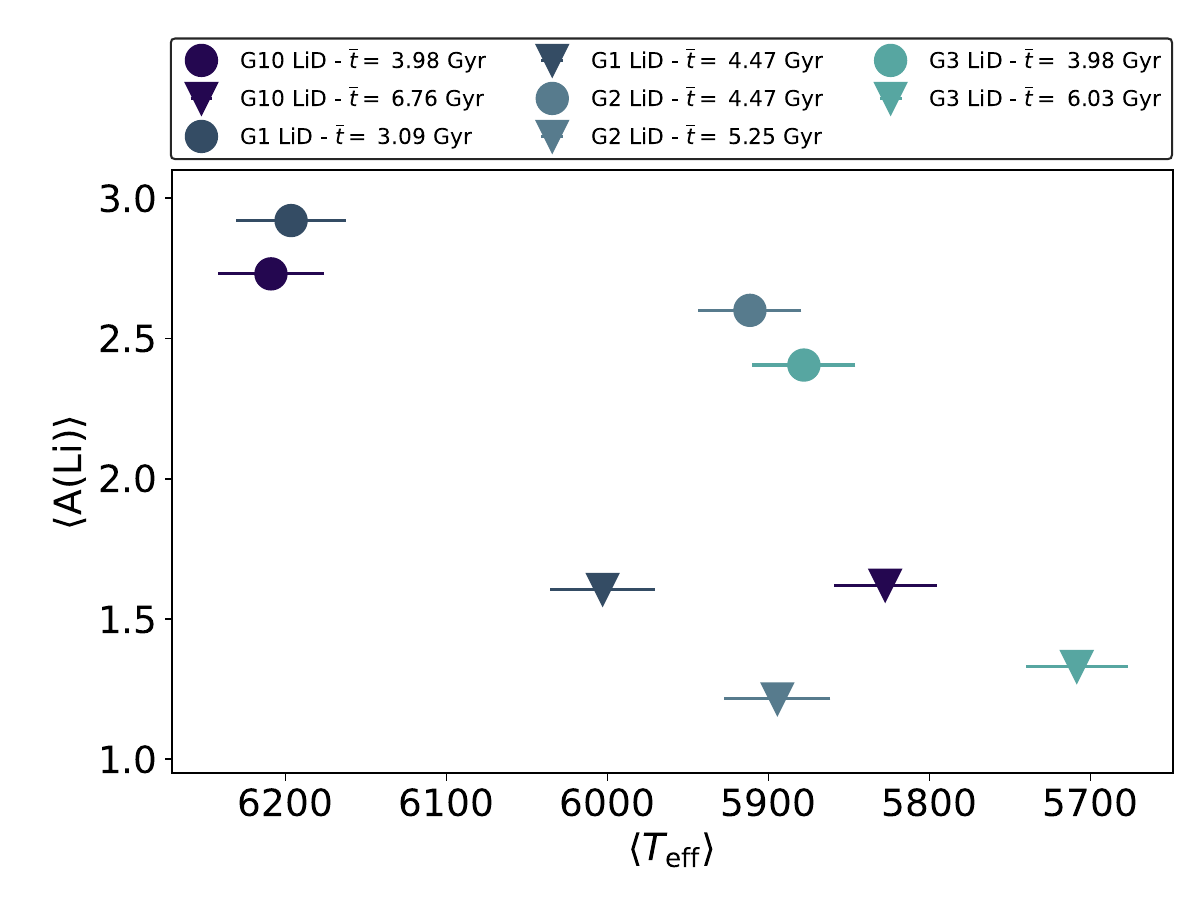}
    \caption{\emph{Left panel:} $\langle {\rm A(Li)} \rangle$ vs \feh\ for the super-solar groups of the sample, split into those with a direct detection of Li (LiD) and those with an upper limit estimate (LiUL). In this representation, only the median of the top six stars with the highest A(Li) is shown in each marker. The black star-shaped markers display the data from \citet{Guiglion2016}. \emph{Right panel:} $\langle {\rm A(Li)} \rangle$ vs $\langle T_{\rm eff} \rangle$. It is possible to see that A(Li) seems to decrease with decreasing $T_{\rm eff}$. Warmer temperatures seem to have a protective effect on A(Li) due to their thinner convective layers of the stars.}
    \label{fig:li}
\end{figure}


\acknowledgements
MLLD acknowledges Agencia Nacional de Investigación y Desarollo (ANID), Chile, Fondecyt Postdoctorado Folio 3240344. MLLD also thanks the support from the International Astronomical Union grant for the IAUS395 2024. MLLD and RS acknowledge support from the National Science Centre, Poland, project 2019/34/E/ST9/00133.


\end{document}